%% file: contactPRAletter.tex
\documentclass[aps,prl,twocolumn,showpacs,floatfix,longbibliography,superscriptaddress]{revtex4-2}
\usepackage{graphicx,amsfonts,amssymb,amsmath}
\usepackage{textcomp} 
\usepackage{esint}
\usepackage[english]{babel}
\usepackage{afterpage}
\usepackage{bbold}
\usepackage{soul}  
\usepackage[colorlinks,linktocpage,citecolor=blue,linkcolor=red,urlcolor=blue]{hyperref}
\def\Gamrpa{\Gamma_{\rm RPA}}
\def\Gamloc{\Gamma_{\rm loc}}
\def\Gloc{G_{\rm loc}}
\def\nloc{n_{\rm MI}}
\def\Zqp{Z^\alpha_{\rm QP}}
\def\Zqpp{Z^+_{\rm QP}}

\def\Cdbg{C_{\rm DBG}}
\def\Pmi{P_{\rm MI}}

\def\Psing{P_{\rm sing}}
\def\nksing{n^{\rm sing}_\k}

\def\nkmi{n_\k^{\rm MI}}
\def\nkbog{n_\k^{\rm Bog}}
\def\nmi{n_{\rm MI}}
\def\cuni{C_{\rm univ}}
\def\mlat{m_{\rm lat}}

\begin{document}
	
\graphicspath{{figures_submit/}}
	
\allowdisplaybreaks
	
\input{./definition.tex}

\title{Two-body contact of a Bose gas near the superfluid--Mott-insulator transition}
	
\author{Moksh Bhateja}
\affiliation{Univ. Lille, CNRS, UMR 8523 – PhLAM – Laboratoire de Physique des Lasers Atomes et Molécules, F-59000 Lille, France}
\affiliation{Sorbonne Universit\'e, CNRS, Laboratoire de Physique Th\'eorique de la Mati\`ere Condens\'ee, LPTMC, F-75005 Paris, France}
	
\author{Nicolas Dupuis}
\affiliation{Sorbonne Universit\'e, CNRS, Laboratoire de Physique Th\'eorique de la Mati\`ere Condens\'ee, LPTMC, F-75005 Paris, France}
	
\author{Adam Ran\c{c}on} 
\affiliation{Univ. Lille, CNRS, UMR 8523 – PhLAM – Laboratoire de Physique des Lasers Atomes et Molécules, F-59000 Lille, France}	
	
\date{September 9, 2025} 
	
\begin{abstract}
The two-body contact is a fundamental quantity of a dilute Bose gas that relates the thermodynamics to the short-distance two-body correlations. For a Bose gas in an optical lattice, near the superfluid--Mott-insulator transition, we show that a ``universal'' contact $\cuni$ can be defined from the singular part $P-\Pmi$ of the pressure ($\Pmi$ is the pressure of the Mott insulator). Its expression $\cuni=C_{\rm DBG}(|n-\nmi|,a^*)$ coincides with that of a dilute Bose gas provided we consider the effective ``scattering length'' $a^*$ of the quasi-particles at the quantum critical point (QCP) rather than the scattering length in vacuum, and the excess density $|n-\nmi|$ of particles (or holes) with respect to the Mott insulator. Close to the transition, we find that the singular part $\nksing = n_\k - \nkmi$ of the momentum distribution exhibits a high-momentum tail of the form $Z_{\rm QP} \cuni/|\k|^4$ over a broad region of the Brillouin zone, where $Z_{\rm QP}$ is the quasi-particle weight of the elementary excitations at the QCP. Our results demonstrate that the notion of contact extends to strongly correlated lattice bosons, and we argue that the contact $\cuni$ can be measured in state-of-the-art experiments on Bose gases in optical lattices and magnetic insulators.  
\end{abstract}

\maketitle

\paragraph{Introduction.} 
 
In a dilute, weakly interacting, Bose gas, the equation of state depends on the atom mass and the $s$-wave scattering length but is otherwise universal, i.e. independent of microscopic details such as the precise shape of the atom-atom interaction potential. Considering the scattering length as an additional thermodynamic variable, in addition to the usual variables (e.g. the chemical potential and the temperature in the grand canonical ensemble), one can define its thermodynamic conjugate, the so-called two-body contact~\cite{Tan08a,Tan08b,Tan08c}. In a dilute gas, the contact relates the (universal) low-temperature thermodynamics to the (universal) short-distance behavior that appears in the two-body correlations or the momentum distribution function~\cite{Tan08a,Tan08b,Tan08c,Olshanii03,Braaten08,Zhang09,Combescot09,Valiente12,Werner12a,Werner12}. To date, few measurements have been made of the contact in Bose gases. In addition to experiments in the thermal regime and quasi-pure Bose-Einstein condensates~\cite{Wild12,Fletcher17,Lopes17}, the contact has been determined in a planar Bose gas in a broad temperature range including normal and superfluid phases~\cite{Zou21,Rancon23}, and in a one-dimensional Lieb-Liniger gas~\cite{Huang24}. 

Strong correlations in a Bose gas can be achieved by loading the gas into an optical lattice. It is then possible, by varying the strength of the lattice potential and/or the density $\bar n$, to induce a quantum phase transition between a superfluid (SF) state and a Mott insulator (MI) where the mean number of bosons per unit cell, $n=\bar n\ell^3$ (with $\ell$ the lattice spacing), is integer~\cite{Greiner02}. When the phase transition is induced by a density change, it belongs to the dilute-Bose-gas universality class~\cite{Fisher89,Sachdev_book,NDbook1}, i.e., it is similar to the quantum phase transition between the vacuum state and the superfluid state obtained by varying the chemical potential from negative to positive values in a dilute gas. This property underpins most of our understanding of the MI-SF transition. In three dimensions, the transition (when induced by a density change) is mean-field-like with a correlation-length exponent $\nu=1/2$ and a dynamical critical exponent $z=2$. Elementary excitations at the quantum critical point (QCP) are quasi-particles (or quasi-holes) that have many similarities with bosons in the absence of the optical lattice: Their dispersion law $E_\k=\k^2/2m^*$ is quadratic in the low-energy limit, with an effective mass $m^*$, and their mutual interaction is determined by an effective ``scattering length'' $a^*$~\cite{Rancon12a,Rancon12d}. In the superfluid phase near the QCP, the system behaves as a dilute gas of weakly interacting quasi-particles (or quasi-holes) with density $|n-\nmi|$ where $\nmi$ is the (integer) density of the Mott insulator. The singular part $\Psing=P-\Pmi$ of the pressure ---that is, the part that is singular when crossing the MI-SF transition by varying the chemical potential or the density--- takes the familiar Bogoliubov form. It includes the Lee-Yang-Huang correction~\cite{Lee57b,Dalfovo99,NDbook1} in addition to the mean-field result, but with the effective mass $m^*$ and the effective scattering length $a^*$ replacing the bare boson mass and the scattering length in vacuum~\cite{Rancon12d}.

In a dilute Bose gas loaded into an optical lattice, short-distance two-body correlations (on scales smaller than the interparticle distance $d\sim \bar n^{-1/3}$) and the large-$\k$ behavior of the momentum distribution $n_\k$ are governed by Tan's relations~\cite{Tan08a,Tan08b,Tan08c}, with a contact whose value is modified by the periodic lattice potential. Near the Mott transition, the filling \(n=\bar n\,\ell^{3}\) is close to an integer, and the contact theory does not apply on scales comparable to or larger than the lattice spacing $\ell\gtrsim d$. However, one may ask whether a ``contact'' can be defined that, by analogy with the dilute gas, controls $n_{\mathbf{k}}$ at high momenta within the first Brillouin zone of the periodic lattice (i.e., $|\mathbf{k}|\lesssim \pi/\ell$). Because $n_{\mathbf{k}}$ generally contains a smooth background that remains finite as $\k$ approaches the zone boundary ---for example, deep in the Mott insulator $n_{\mathbf{k}}\simeq \mathrm{const}$ for $\mathbf{k}\in[-\pi/\ell,\pi/\ell]^3$~\cite{Sengupta05}--- $n_{\mathbf{k}}$ does not exhibit a universal $1/|\mathbf{k}|^{4}$ tail for $|\mathbf{k}|\lesssim \pi/\ell$. Consequently, the ``contact'', if definable below $\pi/\ell$, cannot be extracted from the asymptotic form of $n_{\mathbf{k}}$.

In this Letter, from a strong-coupling random-phase approximation (RPA)~\cite{Sheshadri93,Oosten01,Sengupta05,Ohashi06,Menotti08,Freericks09,Teichmann09a,Teichmann09b,Wang18,Kuebler19,SantAna19,[{For an in-depth presentation of the strong-coupling RPA theory and the detailed study behind the present Letter, see the companion paper: }]Dupuis25} in the Bose-Hubbard model, we show that a ``universal'' two-body contact can be defined from the singular part of the pressure~\cite{[{For another example of a quantum system, which is not a dilute gas but where a contact can be defined, see }]Hofmann23,[{For a general discussion of contact near phase transitions, see }]Chen14} in the superfluid phase near the Mott transition. Its expression, $\cuni=\Cdbg(|n-\nmi|,a^*)$, is the same as in a dilute Bose gas provided we consider the effective scattering length $a^*$ of the quasi-particles at the QCP and the excess density $|n-\nmi|$ of particles (or holes) with respect to the Mott insulator. We also determine the singular part $\nksing=n_\k-\nkmi$ of the momentum distribution, where $\nkmi$ is the distribution in the Mott insulator. We find that $\nksing$ is well described by $Z_{\rm QP} n_\k^{\rm Bog}$, where $n_\k^{\rm Bog}$ is the Bogoliubov result expressed in terms of the effective mass $m^*$ and the distance $|\mu-\mu_c|$ to the QCP ($\mu_c$ is the critical value of the chemical potential at the transition), and $Z_{\rm QP}$ is the quasi-particle weight of the elementary excitations at the QCP. Sufficiently close to the QCP, there is a broad momentum range in the Brillouin zone where the momentum distribution $\nksing$ exhibits the high-momentum tail $Z_{\rm QP}\cuni/|\k|^4$, as in a dilute Bose gas but with an additional prefactor given by $Z_{\rm QP}$.

\paragraph{Strong-coupling RPA.}

The Bose-Hubbard model describes bosons moving on a lattice. The (grand canonical) Hamiltonian is defined by
\beq 
\hat H = \sum_{\r,\r'} t_{\r,\r'} \hat\psi^\dagger_\r \hat\psi_{\r'} + \sum_\r \Bigl( - \mu\hat\psi^\dagger_\r \hat\psi_\r + \frac{U}{2} \hat\psi^\dagger_\r  \hat\psi^\dagger_\r  \hat\psi_\r \hat\psi_\r \Bigr) ,  
\eeq 
where $\hat\psi_\r$ and $\hat\psi^\dagger_\r$ are annihilation and creation operators and the discrete variable $\r$ labels the different sites of a cubic lattice. The hopping matrix is defined by $t_{\r,\r'}=-t$ if $\r$ and $\r'$ are nearest neighbors and $t_{\r,\r'}=0$ otherwise. We denote by $U$ the on-site repulsion between bosons and set the lattice spacing $\ell$ to unity (so that we do not distinguish between the total number of sites $N$ and the volume $\calV=N\ell^3$). The mean boson density $n$ is fixed by the chemical potential $\mu$. 

The ground state of the system can be characterized by the superfluid order parameter $\phi_\r(\tau)=\mean{\hat\psi_\r(\tau)}$ and the boson propagator $G(\r-\r',\tau-\tau')=-\mean{T_\tau \hat\psi_\r(\tau) \hat\psi^\dagger_{\r'}(\tau')}$, where $\hat\psi_\r(\tau)=e^{\hat H\tau }\hat\psi_\r e^{-\hat H\tau}$ is the boson operator in the Heisenberg representation, $\tau\in[0,\beta]$ is an imaginary time with $\beta=1/T\to\infty$ the inverse temperature (we set $\hbar=k_B=1$ throughout), and $T_\tau$ is a time-ordering operator. These quantities can be obtained from the Gibbs free energy 
(which will be referred to as the effective action following the field theory terminology). Considering the hopping term at the mean-field level, while treating exactly the local (on-site) correlations, one obtains the strong-coupling RPA effective action
\begin{multline}
\Gamrpa[\phi^*,\phi] = \Gamloc[0,0] + \inttau d\tau' \sum_{\r,\r'} \phi_\r^*(\tau) \bigl[ t_{\r,\r'} \delta(\tau-\tau') \\ - \delta_{\r,\r'} \Gloc^{-1}(\tau-\tau') \bigr] \phi_{\r'}(\tau') + \frac{g}{2} \inttau \sum_\r |\phi_\r|^4 ,
\label{gamrpa}
\end{multline}
where $\Gloc$ is the (single-site) propagator and $g$ the effective interaction in the local limit (see Appendix). The value of the order parameter $\phi_\r(\tau)=\phi_0$ is obtained by minimizing the effective action.  A nonzero value of the condensate density $n_0=|\phi_0|^2$ implies spontaneous breaking of the U(1) symmetry and superfluidity, while the Mott insulator corresponds to $n_0=0$. 

\begin{figure}
	\centerline{\includegraphics[width=7cm]{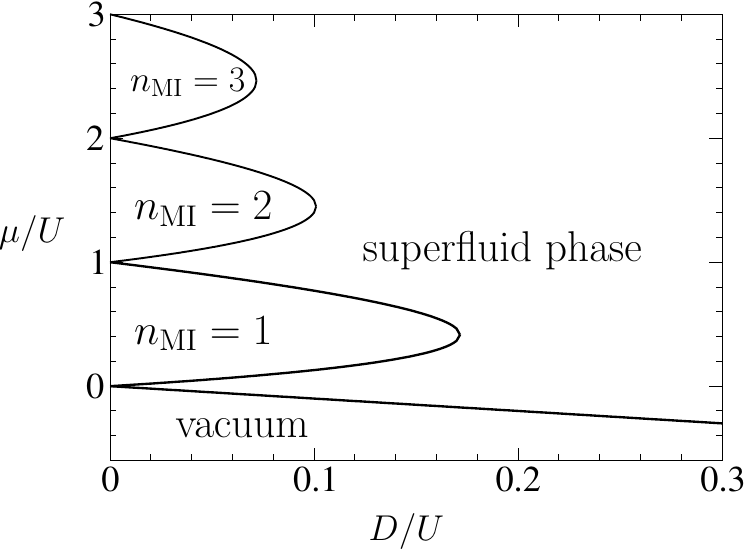}}
	\caption{Phase diagram of the three-dimensional Bose-Hubbard model obtained from the criterion $\Gloc^{-1}(i\wn=0)+D=0$ ($D=6t$). Each Mott lobe is labeled by the integer $\nloc$ giving the mean number of bosons per site.} 
	\label{fig_phase_dia} 
\end{figure}

\paragraph{Mott insulator and MI-SF transition.} 

The MI-SF transition corresponds to a change of sign of the quadratic term in the effective action, i.e. $\Gloc^{-1}(i\wn=0)+D=0$, where $D=-t_{\k=0}=6t$. 
In the plane $(D/U,\mu/U)$, the phase diagram is given by a series of Mott lobes $\mu_-(\nloc)\leq \mu\leq \mu_+(\nloc)$, labeled by the integer (mean) number of bosons per site $\nloc\equiv \nloc(\mu)$, as shown in Fig.~\ref{fig_phase_dia}, in agreement with previous mean-field studies~\cite{Sheshadri93,Oosten01,Sengupta05,Kopec24}. The position of the tip of the Mott lobes is defined by $D_c/U=2\nloc+1-2(\nloc^2+\nloc)^{1/2}$ and $\mu_c=U(\nloc-1/2)-D_c/2$. Since $\Gamrpa[0,0]=\Gamloc[0,0]$, a straightforward calculation gives the pressure $P_{\rm MI}=\mu\nloc-(U/2)\nloc(\nloc-1)$ in the Mott insulator. The mean density $n=\partial P_{\rm MI}/\partial\mu=\nloc$ is constant and the compressibility $\kappa=\partial n/\partial \mu$ vanishes. When $\mu$ becomes larger (smaller) than $\mu_+$ ($\mu_-$), the density deviates from $\nmi$ and the system becomes superfluid. 

\begin{figure}
	\centerline{\includegraphics[width=4cm]{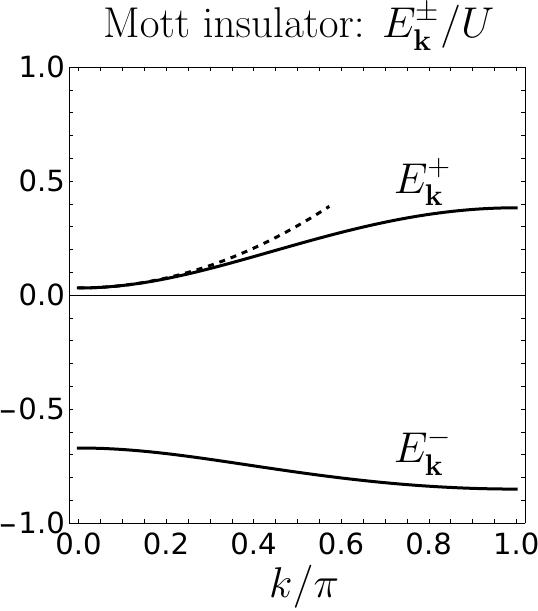}
		\hspace{0.3cm}
		\includegraphics[width=4cm]{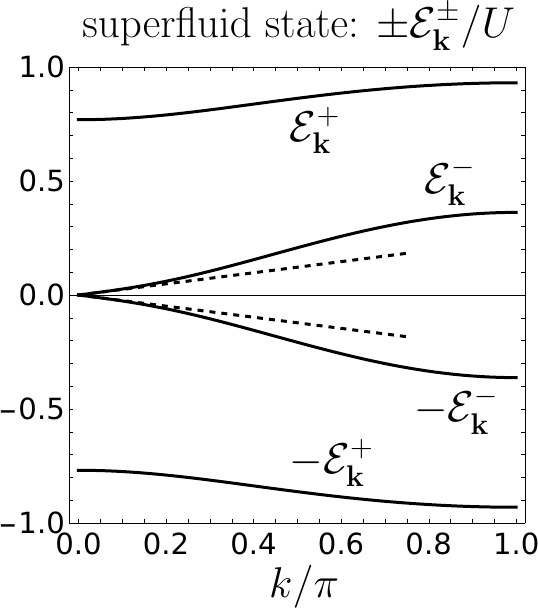}}
	\caption{Excitation energies in the Mott insulator ($\mu=0.9\,\mu_+$, left) and in the superfluid state ($\mu=1.1\,\mu_+$, right), along the Brillouin zone diagonal $\k=(k,k,k)$, for $\nloc=1$. The dashed lines show the approximate low-energy forms, valid near $\k=0$, $E_\k^+= \k^2/2m^*_+ + \mu_+ - \mu$ and $\calE_\k^-=c|\k|$ (with $c$ the sound velocity in the superfluid state).}
	\label{fig_spectrum} 
\end{figure}

\begin{figure}
	\centerline{\includegraphics[width=4cm]{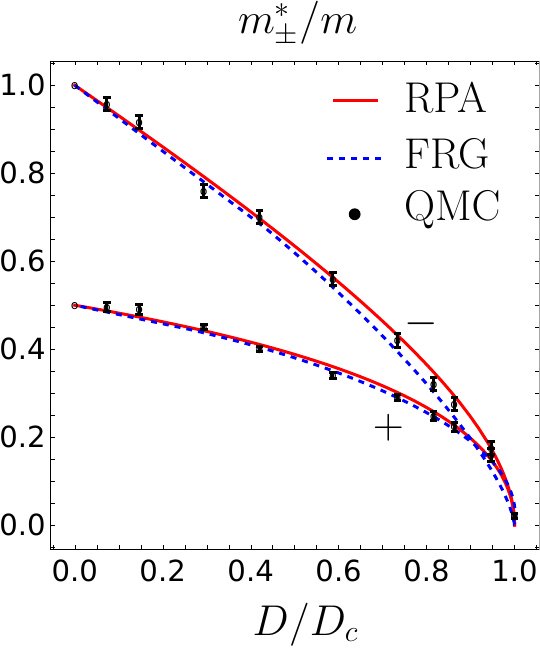}
		\hspace{0.3cm}
		\includegraphics[width=4cm]{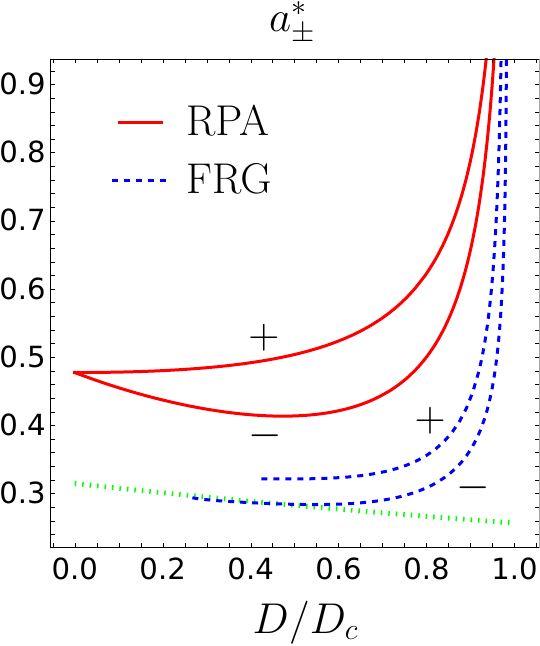}}
	\caption{Effective mass $m^*_\alpha$ (left) and effective scattering length $a^*_\alpha$ (right) vs $D/D_c$ at the quantum critical point between the Mott insulator $\nmi=1$ and the superfluid state, obtained from strong-coupling RPA, nonperturbative functional renormalization group (FRG)~\cite{Rancon12d} and quantum Monte Carlo simulations (QMC)~\cite{Capogrosso07}. The green dotted line in the right panel shows the (vacuum) scattering length $a_{\rm lat}\simeq 1/[8\pi(t/U+0.1264)]$ of the bosons moving on the lattice~\cite{Rancon11b}. The strong-coupling RPA is reliable for the effective mass but less so for the effective scattering length, even if the general trend is correct.}
	\label{fig_mastar}.  
\end{figure}

Near the MI-SF transition, i.e. when $\mu$ is close to $\mu_\alpha$ (with $\alpha=\pm$), the propagator takes the quasi-particle form 
\beq 
G(\k,i\wn) \simeq \frac{\alpha \Zqp}{i\wn - E^\alpha_\k} 
\eeq 
in the low-energy limit. The one-particle excitation spectrum $E^\alpha_\k=\alpha(\k^2/2m^*_\alpha + \Delta_\alpha)$ is particle-like for $\mu$ near $\mu_+$ and hole-like for $\mu$ near $\mu_-$ (Fig.~\ref{fig_spectrum}).  The excitation gap $\Delta_\alpha=\alpha(\mu_\alpha-\mu)$ vanishes linearly with $\mu_\alpha-\mu$, implying $z\nu=1$. At the QCP ($\mu=\mu_\alpha$), $E^\alpha_\k=\alpha \k^2/2m^*_\alpha$ so that $z=2$ and $\nu=1/2$. The spectral weight $\Zqp$ and the effective mass $m^*_\alpha$ (both positive) of the quasi-particles are defined by
\beq 
\Zqp = \frac{\mlat}{m^*_\alpha} = \alpha \frac{\Gloc(0)}{D\Gloc'(0)} ,
\eeq 
where $\mlat=1/2t$ is the effective mass of the free bosons moving on the cubic lattice. It is natural to introduce the quasi-particle interaction strength $g^\alpha_R=g(\Zqp)^2$. By analogy with the dilute Bose gas, we can then define an effective ``scattering length'' $a^*_\alpha$ by $g^\alpha_R|_{\mu=\mu_\alpha} = 4\pi a^*_\alpha/m^*_\alpha$~\cite{Rancon12d} (Fig.~\ref{fig_mastar}).

\paragraph{Superfluid phase and definition of the contact.} 
  
In the superfluid state, the condensate density $n_0=\alpha(\mu-\mu_\alpha)/\Zqp g$ is nonzero and the pressure is given by $P= \Pmi+(\mu-\mu_\alpha)^2/2\Zqp g$, i.e., 
\beq 
P = \Pmi + \frac{m^*_\alpha}{8\pi a^*_\alpha} (\mu-\mu_\alpha)^2 .
\label{Pscal}
\eeq 
The singular part $\Psing=P-\Pmi$ of the pressure exhibits the standard (mean-field) Bogoliubov form but with the effective mass $m^*_\alpha$ and the effective scattering length $a^*_\alpha$ instead of the bare boson mass and scattering length in vacuum, and the distance $|\mu-\mu_\alpha|$ to the QCP rather than the chemical potential. Note that $\Psing$ is independent of the quasi-particle weight.
The mean density 
\beq 
n = \frac{\partial P}{\partial \mu} = \nloc+\frac{m^*_\alpha}{4\pi a^*_\alpha}(\mu-\mu_\alpha)
\eeq 
and the condensate density 
\beq 
n_0 = \Zqp \frac{m^*_\alpha}{4\pi a^*_\alpha} |\mu-\mu_\alpha| = \Zqp |n-\nmi|
\label{n0}
\eeq 
can also be expressed in terms of $m^*_\alpha$ and $a^*_\alpha$, whereas the superfluid density $n_s$ is equal to $|n-\nmi|$~\cite{Dupuis25}. The dependence of $n_0$ on the quasi-particle weight $\Zqp$ is due to the Bose-Einstein condensation involving not particles but quasi-particles (or quasi-holes). Since $\Zqp=m/m^*_\alpha\geq 1$ (Fig.~\ref{fig_mastar}), $n_0$ is larger than $|n-\nmi|$: The excess particles (holes) with respect to the Mott insulator drag other particles (holes) into the condensation. 

We can now define a universal contact in the usual way, i.e. by taking the derivative of the singular part of the pressure with respect to the effective scattering length $a^*_\alpha$,
\begin{align}
\cuni &= 8\pi m^*_\alpha \calV \frac{\partial \Psing(\mu-\mu_\alpha,m^*_\alpha,a^*_\alpha)}{\partial (1/a^*_\alpha)} \biggl|_{\mu-\mu_\alpha,m^*_\alpha} \nonumber\\ 
&= \calV [ m^*_\alpha(\mu-\mu_\alpha)]^2 ,
\label{Cmu}
\end{align}  
which is analog to the result of Bogoliubov's theory for a dilute Bose gas, but with the effective mass $m^*_\alpha$ instead of the bare boson mass. Alternatively, one can express the contact as a function of the mean density, 
\beq 
\cuni = \calV [ 4\pi a^*_\alpha(n - \nloc)]^2 . 
\label{Cn}
\eeq 
We recover the Bogoliubov expression of the contact of superfluid bosons with density $|n-\nloc|$ and scattering length $a^*_\alpha$.

\paragraph{Momentum distribution.}

In the superfluid state, the two bands $E^\pm_\k$ of the Mott insulator split into four bands
$\pm\calE_\k^\pm$ as shown in Fig.~\ref{fig_spectrum}~\cite{Sengupta05,Ohashi06,Huber07,Menotti08}. Consider the case of particle doping ($\mu>\mu_+$ and $n>\nmi$) where the positive energy band $E_\k^+$ of the Mott insulator becomes gapless (similar results are obtained in the case of hole doping). In that case, the band $E_\k^-$ evolves into the band $-\calE_\k^+$ and the band $E_\k^+$ into $\calE_\k^-$. Two new bands, $\calE_\k^+$ and $-\calE_\k^-$, appear in the superfluid. The band $\calE_\k^+$ carries a negligible fraction of the spectral weight in the vicinity of the transition. 

Ignoring the contribution $\delta_{\k,0}Nn_0$ of the condensate, the singular part $\nksing = n_\k-\nkmi$ of the momentum distribution can be written as
\beq 
\nksing = - \calS(-\calE_\k^-) - \calS(-\calE_\k^+) + \calS_{\rm MI}(E_\k^-) ,
\label{nksing}  
\eeq 
where $\nkmi$ is the ($\mu$-independent) distribution in the Mott insulator and $\calS(-\calE_\k^\pm)$ ($\calS_{\rm MI}(E_\k^-)$) denotes the spectral weight associated with the energy $-\calE_\k^\pm$ ($E_\k^-$) of the excitations in the superfluid (Mott insulator)~\cite{not11}. The gapped band is little affected when $\mu$ becomes larger than $\mu_+$ and $\calS(-\calE_\k^+)$ is essentially equal to $\calS_{\rm MI}(E_\k^-)$ near the transition. On the other hand, although the band $\calE_\k^-$ carries most of the spectral weight of the band $E_\k^+$ of the Mott insulator, the gapless negative energy band $-\calE^-_\k$ gives a large contribution to the momentum distribution for small momenta, as in a dilute superfluid gas. This implies that $\nksing$ is well approximated by $-\calS(-\calE^-_\k)$. Furthermore, we find that $-\calS(-\calE^-_\k) \simeq \Zqpp \nkbog$, where
\beq 
\nkbog = - \half + \frac{\eps_\k+\mu-\mu_+}{2\sqrt{\eps_\k[\eps_\k + 2(\mu-\mu_+)]}} 
\label{nkbog} 
\eeq 
is the standard Bogoliubov result for bosons (ignoring the contribution of the condensate) with dispersion $\eps_\k=\k^2/2m^*_+$ and chemical potential $\mu-\mu_+$. When $|\k|$ is larger than the characteristic momentum scale $k_*=2[m^*_+(\mu-\mu_+)]^{1/2}$, $n^{\rm Bog}_{\k}\simeq \cuni/\calV |\k|^4$ with the contact $\cuni$ defined by~(\ref{Cmu}) so that the momentum distribution~\cite{not2}, 
\beq 
\nksing \simeq \frac{\Zqpp \cuni}{\calV|\k|^4} \qquad (k_*\ll |\k|) ,
\eeq 
exhibits a high-momentum tail $\sim 1/|\k|^4$, as in a dilute Bose gas, provided the characteristic scale $k_*$ is small enough. This requires the system to be sufficiently close to the QCP.  

\begin{figure}
	\centerline{\includegraphics[width=3.95cm]{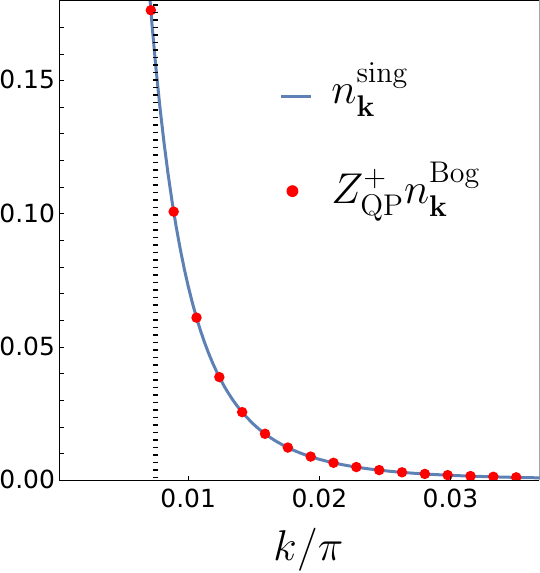}
		\includegraphics[width=4cm]{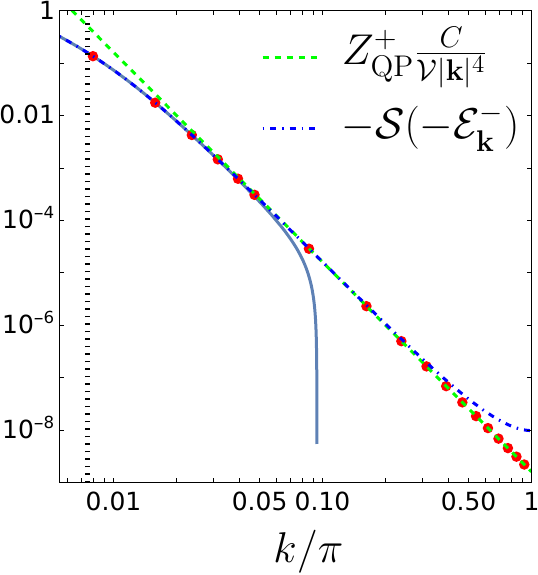}
	}
	\caption{Singular part $\nksing=n_\k-n_\k^{\rm MI}$ of the momentum distribution, along the Brillouin zone diagonal $\k=(k,k,k)$, for $D=D_c/2$ and $\mu=1.000005\mu_+$ ($n=1.0003$), compared to $-\calS(-\calE_\k^-)$ (spectral weight of the negative gapless energy branch), $Z^+_{\rm QP}\nkbog$~[Eq.~(\ref{nkbog})] and $Z_{\rm QP}^+ \cuni/\calV|\k|^4$ where $\cuni$ is the contact defined from the pressure [Eq.~(\ref{Cmu})]. The vertical dotted line shows the momentum scale $k_*/\sqrt{3}$ where $k_*=2(m^*_\alpha|\mu-\mu_\alpha|)^{1/2}$.}
	\label{fig_momentum} 
\end{figure}

Figure~\ref{fig_momentum} shows the momentum distribution obtained from Eq.~(\ref{nksing}). The relation $-\calS(-\calE^-_\k)\simeq \Zqpp \nkbog$ holds over the entire Brillouin zone ---except very close to the zone boundaries where the free dispersion $t_\k$ differs from $\k^2/2\mlat-D$ due to lattice effects--- but the agreement between $\nksing$ and $-\calS(-\calE_\k^-)$ breaks down when $|\k|\gtrsim 0.25$ for $\k$ varying along the Brillouin zone diagonal. We believe that the high-momentum tail $\Zqpp \cuni/\calV|\k|^4$ should be observed up to the Brillouin zone boundaries. The momentum sum rule $n=n_0+(2\pi)^{-3}\int d^3k\, n_\k$ is only satisfied to within $10^{-4}$ in the case of Fig.~\ref{fig_momentum}. The slight difference between $-\calS(-\calE_\k^+)$ and $\calS_{\rm MI}(E^-_\k)$ at large momenta, which is also of the order $10^{-4}$ and spoils the agreement between $\nksing$ and $-\calS(-\calE_\k^-)$, is thus likely to be an artifact of the strong-coupling RPA. This expectation is confirmed by a study of a hard-core boson model ---which should describe the hole-doped Mott insulator $\nmi=1$ in the limit $t/U\ll 1$--- where the high-momentum tail $\Zqpp \cuni/\calV|\k|^4$ is indeed observed up to the Brillouin zone boundaries (see Fig.~6 in~\cite{Dupuis25}).

The high-momentum tail of $\nksing$ depends not only on the contact but also on the quasi-particle weight $\Zqp$. As pointed out above, this is due to the role of quasi-particles (or quasi-holes) in the condensation leading to superfluidity. If we introduce the quasiparticle field $\bar\psi_\r(\tau)=(\Zqp)^{-1/2}\psi_\r(\tau)$, one finds that the quasi-particle condensate density $\bar n_0=n_0/\Zqp\equiv \bar n_0(|\mu-\mu_\alpha|,m^*_\alpha)$ takes the usual Bogoliubov mean-field expression~\cite{Rancon12d}, and the singular part of the momentum distribution $\bar n_\k^{\rm sing}\simeq\nkbog$ is independent of $\Zqp$.

\paragraph{Conclusion.} 

The fact that the SF-MI transition belongs to the dilute-Bose-gas universality class not only determines the mean-field-like behavior of the transition. It also implies that the superfluid phase, in the vicinity of the transition occurring at $\mu=\mu_+$ or $\mu=\mu_-$, is well described by Bogoliubov's theory, provided we consider the excess of particles (or holes) with respect to the Mott insulator. The elementary excitations at the QCP are quasi-particles with effective mass $m^*_\alpha$, effective scattering length $a^*_\alpha$, and spectral weight $\Zqp$ ($\alpha=\pm$). This allows us to define, from the singular part of the pressure in the superfluid state, a universal two-body contact $\cuni$ which takes the usual Bogoliubov form when expressed in terms of $|n-\nmi|$ and $a^*_\alpha$. Remarkably, there is a broad momentum range in the Brillouin zone where the singular part $\nksing=n_\k-\nkmi$ of the momentum distribution exhibits a high-momentum tail $\Zqp \cuni/\calV|\k|^4$, as in a dilute Bose gas, but with the quasi-particle weight as an additional prefactor. This description of a strongly interacting superfluid system has similarities with the case of a doped fermionic Mott insulator, described by Fermi liquid theory (and therefore by a few effective parameters such as the quasi-particle effective mass and the Landau parameters) like any other conventional metal (strongly or weakly interacting).  

Single-atom-resolved measurement of the momentum distribution~\cite{Chang16,Cayla18,Cayla23} in a Bose gas loaded into an optical lattice should allow the determination of $n_\k$ in the vicinity of the superfluid--Mott-insulator transition. Assuming a flat-box potential, where the density $n$ is controlled, the singular part $\nksing$ of the momentum distribution can be determined by tuning the gas across the transition. Observation of the $1/|\k|^4$ tail then gives $Z_{\rm QP}\cuni$. The condensate density $n_0$ being directly obtained from $n_{\k=0}$, we deduce the effective scattering length $a^*$ and the quasi-particle weight $Z_{\rm QP}$. A fit of $\nksing$ to the Bogoliubov form~(\ref{nkbog}) would provide us with the value of the effective mass $m^*$. Alternatively, $m^*$ can be obtained by measuring the pressure [Eq.~(\ref{Pscal})] or the superfluid transition temperature~\cite{Rancon12d}. On the other hand, the momentum distribution of a Bose gas in an optical lattice can also be measured in the high-momentum range $|\k|\gg 1/\ell$ (a range outside the Bose-Hubbard model). At these short length scales, the system behaves as a dilute gas and we expect a $1/|\k|^4$ tail with a strength given by the ``full'' contact $C$, related to the total pressure of the gas~\cite{Dupuis25}.  

Similarly to the phenomenon of superfluidity, Bose-Einstein condensation of magnons occurs in magnetic insulators~\cite{Giamarchi08}. In the simplest cases, these systems are effectively described by the Bose-Hubbard model in the hard-core limit, with the applied external magnetic field playing the role of the chemical potential. The paramagnetic state is analogous to the Mott insulator, while the magnetically ordered state corresponds to the superfluid state, and the quantum phase transition between the two belongs to the dilute-Bose-gas universality class~\cite{Rancon14a}. The spin structure factor plays the role of momentum distribution; the $1/|\k|^4$ tail can be measured by inelastic neutron scattering, whereas $m^*$ can be obtained from the critical temperature. Therefore, universal contact, as well as effective scattering length $a^*$ and effective mass $m^*$, could be measured in magnetic insulators.

We thank D. Cl\'ement and T. Chalopin for useful discussions and comments, and B. Capogrosso-Sansone for providing us with the QMC data~\cite{Capogrosso07} shown in Fig.~\ref{fig_mastar}. MB was supported by a Charpak fellowship and the Labex CEMPI (ANR-11-LABX-0007-01). AR is supported in part by an IEA CNRS project, and by the “PHC COGITO” program (project number: 49149VE) funded by the French Ministry for Europe and Foreign Affairs, the French Ministry for Higher Education and Research, and the Croatian Ministry of Science and Education.

\input{./contactPRL_final.bbl}

\end{document}

%% file: definition.tex

\newcommand{\oldnew}[2]{\marginpar{\scriptsize \textcolor{red}{correction}}{\textcolor{red}{#2}}}
\newcommand{\suppressed}[1]{\marginpar{\scriptsize \textcolor{red}{correction}}{\textcolor{red}{\st{#1}}}}
\newcommand{\correction}[1]{\marginpar{\textcolor{red}{\scriptsize #1}}}
\newcommand{\marge}[1]{\marginpar{\scriptsize #1}}
\newcommand{\remarque}[1]{\marginpar{\scriptsize Remarque}{\it [#1]}}

%

\def\rhoeq{\hat\rho_{\rm eq}}

\newcommand{\new}[1]{{\bf #1}}
\newlength{\textlarg}
\newcommand{\redbar}[1]{\textcolor{red}{\st{#1}}} 
\newcommand{\bluebar}[1]{\textcolor{blue}{\st{#1}}} 

\newcommand{\normord}[1]{:\mathrel{#1}:}

\newcommand{\beq}{\begin{equation}}
\newcommand{\eeq}{\end{equation}}
\newcommand{\bfig}{\begin{figure}}
\newcommand{\efig}{\end{figure}}
\newcommand{\bline}{\begin{multline}}
\newcommand{\eline}{\end{multline}}
\newcommand{\bremark}{\begin{quotation} \noindent \small }
\newcommand{\eremark}{\end{quotation}}
\newcommand{\llbrace}{\left\lbrace}  
\newcommand{\rrbrace}{\right\rbrace}
\newcommand{\lbraket}{\left[}
\newcommand{\rbraket}{\right]}
\newcommand{\llangle}{\left\langle}
\newcommand{\rrangle}{\right\rangle} 

\newcommand{\Tr}{{\rm Tr}} 
\newcommand{\tr}{{\rm tr}} 
\newcommand{\sgn}{\,{\rm sgn}} 
\newcommand{\mean}[1]{\langle #1 \rangle}
\newcommand{\commu}[2]{[#1,#2]} 
\newcommand{\bra}[1]{\langle#1|}
\newcommand{\ket}[1]{|#1\rangle}
\newcommand{\braket}[2]{\langle #1|#2\rangle}
\newcommand{\ketbra}[2]{|#1\rangle\langle#2|}
\newcommand{\dbraket}[3]{\langle #1|#2|#3\rangle}
\newcommand{\tens}[1]{\overleftrightarrow{#1}}  
\newcommand{\vac}{|{\rm vac}\rangle} 
\newcommand{\bravac}{\langle{\rm vac}|}
\newcommand{\const}{{\rm const}} 
\newcommand{\unif}{{\rm unif.}} 
\newcommand{\atanh}{\,{\rm atanh}}
\newcommand{\cotanh}{\,{\rm cotanh}}

\newcommand{\ie}{i.e.\xspace}
\newcommand{\iet}{i.e.}
\newcommand{\eg}{e.g.\xspace}
\newcommand{\cc}{{\rm c.c.}} 
\newcommand{\hc}{{\rm H.c.}} 
\newcommand{\etal}{{\it et al. }}
\newcommand\eme{$^{\mbox{\footnotesize ème}}$\xspace}

\newcommand{\jhatbf}{\hat {\textbf \jold}} 
\newcommand{\Jhatbf}{\hat {\textbf \J}} 
\newcommand{\jhat}{\hat {\jmath}} 
\newcommand{\Jhat}{\hat {J}} 
\newcommand{\jbf}{\textbf j}
\newcommand{\Jbf}{\textbf J}

\def\chibf{\boldsymbol{\chi}}
\def\down{\downarrow}
\def\eps{\epsilon}
\def\gam{\gamma} 
\def\alphabf{\boldsymbol{\alpha}}
\def\phibf{\boldsymbol{\phi}}
\def\varphibf{\boldsymbol{\varphi}}
\def\varphibfs{\boldsymbol{\varphi}_<}
\def\varphibfl{\boldsymbol{\varphi}_>}
\def\varphis{\varphi_{<}}
\def\varphil{\varphi_{>}}
\def\psibf{\boldsymbol{\psi}}
\def\thetabf{\boldsymbol{\theta}}
\def\Ome{\Omega}
\def\omeD{{\omega_D}} 
\def\bfOme{\boldsymbol{\Omega}} 
\def\Omebf{\boldsymbol{\Omega}} 
\def\lamb{\lambda}
\def\Lamb{\Lambda}
\def\sig{\sigma}
\def\Sig{\Sigma}
\def\sigp{{\sigma'}} 
\def\bfsig{\boldsymbol{\sigma}} 
\def\sigbf{\boldsymbol{\sigma}} 
\def\bfSig{\boldsymbol{\Sigma}} 
\def\The{\Theta} 
\def\up{\uparrow}

\def\epsk{\epsilon_{\bf k}} 
\def\epsp{\epsilon_{\bf p}} 
\def\xik{\xi_{\bf k}} 
\def\txik{\tilde\xi_{\bf k}} 
\def\xip{\xi_{\bf p}} 
\def\epsq{\epsilon_{\bf q}} 
\def\xiq{\xi_{\bf q}} 
\def\xikq{\xi_{{\bf k}+{\bf q}}} 
\def\Ek{E_{\bf k}} 
\def\Ep{E_{\bf p}}
\def\Eq{E_{\bf q}}
\def\Heff{\hat H_{\rm eff}}
\def\Hem{\hat H_{\rm em}}
\def\Hint{\hat H_{\rm int}}
\def\Hloc{\hat H_{\rm loc}}
\def\HMF{\hat H_{\rm MF}}
\def\HLL{\hat H_{\rm LL}}
\def\Hdis{\hat H_{\rm dis}}
\def\Sem{S_{\rm em}}
\def\SMF{S_{\rm MF}} 
\def\SHF{S_{\rm HF}} 
\def\SRPA{S_{\rm RPA}} 
\def\Sint{S_{\rm int}} 
\def\Sloc{S_{\rm loc}}
\def\TN{T_{\rm N}} 
\def\TNHF{T^{\rm HF}_{\rm N}} 
\def\Zloc{Z_{\rm loc}} 
\def\ZMF{Z_{\rm MF}} 
\def\ZHF{Z_{\rm HF}} 
\def\ZRPA{Z_{\rm RPA}} 
\def\RPA{{\rm RPA}}
\def\loc{{\rm loc}} 
\def\pp{{\rm pp}}
\def\ph{{\rm ph}} 
\def\ch{{\rm ch}}
\def\sp{{\rm sp}} 
\def\qtf{q_{\rm TF}}
\def\epstf{\eps^{}_{\rm TF}} 
\def\epsrpa{\eps^{}_{\rm RPA}} 
\def\chinnzpp{\chi_{nn}^{0}{}\!\!\!''}
\def\SigHF{\Sigma_{\rm HF}}
\def\psicl{\psi_{\rm cl}} 

\def\half{\frac{1}{2}}
\def\dhalf{\dfrac{1}{2}}
\def\third{\frac{1}{3}} 
\def\quarter{\frac{1}{4}}

\def\qr{{\bf q}\cdot{\bf r}}
\def\wt{\omega t} 

\def\a{{\bf a}}
\def\b{{\bf b}}
\newcommand{\cv}{{\bf c}} 
\def\e{{\bf e}}
\def\f{{\bf f}}
\def\g{{\bf g}}
\def\h{{\bf h}}
\def\jold{\char"11}
\def\j{{\bf j}}
\def\k{{\bf k}}
\def\l{{\bf l}}
\def\ellbf{\bm{\ell}} 
\def\m{{\bf m}}
\def\n{{\bf n}} 
\def\p{{\bf p}} 
\def\q{{\bf q}}
\def\r{{\bf r}}
\def\t{{\bf t}}
\def\u{{\bf u}}
\newcommand{\vv}{{\bf v}}
\def\x{{\bf x}}
\def\y{{\bf y}} 
\def\z{{\bf z}} 
\def\A{{\bf A}}
\def\B{{\bf B}}
\def\D{{\bf D}} 
\def\E{{\bf E}} 
\def\F{{\bf F}} 
\def\H{{\bf H}}  
\def\J{{\bf J}}
\def\K{{\bf K}} 

\def\G{{\bf G}}
\def\L{{\bf L}}
\def\M{{\bf M}}  
\def\O{{\bf O}} 
\def\P{{\bf P}} 
\def\Q{{\bf Q}} 
\def\R{{\bf R}}
\def\S{{\bf S}}
\def\U{{\bf U}} 
\def\V{{\bf V}} 
\def\X{{\bf X}} 
\def\Y{{\bf Y}} 
\def\epsbf{\boldsymbol{\epsilon}}
\def\betabf{\boldsymbol{\beta}}
\def\deltabf{\boldsymbol{\delta}}
\def\mubf{\boldsymbol{\mu}}
\def\nablabf{\boldsymbol{\nabla}}
\def\rhobf{\boldsymbol{\rho}}
\def\sigmabf{\boldsymbol{\sigma}} 
\def\Pibf{\boldsymbol{\Pi}}
\def\pibf{\boldsymbol{\pi}}

\def\para{\parallel}
\def\kpara{{k_\parallel}}
\def\kperp{{k_\perp}} 
\def\kperpp{{k_\perp'}} 
\def\qperp{{q_\perp}} 
\def\tperp{{t_\perp}} 

\def\w{\omega}
\def\wn{\omega_n}
\def\wm{\omega_m}
\def\wnu{\omega_\nu}
\def\wp{\omega_p} 
\def\dmu{{\partial_\mu}}
\def\dnu{{\partial_\nu}}
\def\dl{{\partial_l}}  
\def\dt{\partial_t} 
\def\tdt{\tilde\partial_t}
\def\dk{\partial_k}
\def\tdk{\tilde\partial_k}
\def\dx{\partial_x}
\def\dy{\partial_y} 
\def\dw{\partial_{\w}}
\def\dtau{{\partial_\tau}}  
\def\det{{\rm det}} 
\def\Pf{{\rm Pf}}
\def\diag{{\rm diag}}

\def\dsum{\displaystyle \sum}
\def\dint{\displaystyle \int} 
\def\intt{\int_{-\infty}^\infty dt} 
\def\inttp{\int_{-\infty}^\infty dt'} 
\def\intk{\int_{\bf k}} 
\def\intkd{\int \frac{d^dk}{(2\pi)^d}}
\def\intq{\int_{\bf q}} 
\def\intr{\int d^dr}  
\def\dintr{\displaystyle \int d^dr} 
\def\intrp{\int d^dr'}
\def\dinttau{\displaystyle \int_0^\beta d\tau}
\def\dinttaup{\displaystyle \int_0^\beta d\tau'}
\def\inttau{\int_0^\beta d\tau}
\def\inttaup{\int_0^\beta d\tau'}
\def\intx{\int d^{d+1}x} 
\def\inttaur{\int_0^\beta d\tau \int d^dr}
\def\intinf{\int_{-\infty}^\infty}
\def\dinttaur{\displaystyle \int_0^\beta d\tau \int d^dr}
\def\dintinf{\displaystyle \int_{-\infty}^\infty}
\def\intw{\int_{-\infty}^\infty \frac{d\w}{2\pi}}
\def\sumr{\sum_{\bf r}} 

\def\calA{{\cal A}}
\def\calAbf{\bm{{\cal A}}}
\def\calB{{\cal B}} 
\def\calC{{\cal C}} 
\def\dt{\partial_t}
\def\calD{{\cal D}}
\def\calE{{\cal E}}
\def\calF{{\cal F}} 
\def\calFbf{\bm{{\cal F}}}
\def\calG{{\cal G}}
\def\calH{{\cal H}}
\def\calI{{\cal I}}
\def\calJ{{\cal J}}
\def\calK{{\cal K}}
\def\calL{{\cal L}} 
\def\calM{{\cal M}} 
\def\calN{{\cal N}}
\def\calO{{\cal O}}
\def\calP{{\cal P}}  
\def\calR{{\cal R}} 
\def\calS{{\cal S}}
\def\calT{{\cal T}}
\def\calU{{\cal U}}
\def\calV{{\cal V}}
\def\calX{{\cal X}} 
\def\calY{{\cal Y}} 
\def\calW{{\cal W}} 
\def\calZ{{\cal Z}}

\def\tT{{\tilde T}}
\def\talpha{{\tilde\alpha}}
\def\tbeta{{\tilde\beta}}
\def\tchi{{\tilde\chi}}
\def\tdelta{{\tilde\delta}}
\def\tDelta{{\tilde\Delta}}
\def\teta{{\tilde\eta}} 
\def\tlamb{{\tilde\lambda}}
\def\tmu{{\tilde\mu}}
\def\tphibf{{\tilde\phibf}}
\def\trho{{\tilde\rho}}
\def\tvarphibf{{\tilde\varphibf}} 
\def\tq{\tilde q}
\def\tw{{\tilde\omega}}
\def\twn{{\tilde\omega_n}}
\def\twnu{{\tilde\omega_\nu}}

\def\asinh{{\rm asinh}} 
\def\Tbkt{T_{\rm BKT}}

%% file: contactPRL_final.bbl
%

%% file: contactPRAletter.bbl
\begin{thebibliography}{49}
\bibitem [{not({\natexlab{a}})}]{not11}%
  \BibitemOpen
  \bibinfo {note} {The spectral weight associated with a
  given excitation corresponds to the residue of the pole of the one-particle
  Green function at that energy.}\BibitemShut {Stop}%
\end{thebibliography}
